\documentclass[prl,twocolumn,superscriptaddress,showpacs,floatfix]{revtex4}
\usepackage{epsfig}

\begin{document}

\title{A Limit of Stability in Supercooled Liquid Clusters.}

\author{Eduardo Mendez-Villuendas}
\affiliation{Department of Chemistry, University of Saskatchewan,
Saskatoon, Saskatchewan, S7N 5C9, Canada}

\author{Ivan Saika-Voivod\footnote{Present address: Dept. Physics and Physical Oceanography, Memorial University of Newfoundland, St Johns, NL, A1B 3X7, Canada}}
\affiliation{Department of Chemistry, University of Saskatchewan,
Saskatoon, Saskatchewan, S7N 5C9, Canada}

\author{Richard K. Bowles\footnote{Corresponding author email: richard.bowles@usask.ca}}
\affiliation{Department of Chemistry, University of Saskatchewan,
Saskatoon, Saskatchewan, S7N 5C9, Canada}

\date{\today}
\begin{abstract}
We examine the metastable liquid phase of a supercooled gold nanocluster by studying the free energy landscape using the largest solid-like embryo as an order parameter. Just below freezing, the free energy exhibits a local minimum at small embryo sizes and a maximum at larger embryo sizes which denotes the critical embryo size. At $T=660$K the free energy becomes a monotonically decreasing function of the order parameter as the liquid phase becomes unstable, indicating we have reached a spinodal. In contrast to the usual mean-field theory predictions, the size of the critical embryo remains finite as the spinodal is approached. We also calculate the rate of nucleation, independently from our free energy calculations, and observe a rapid increase in its temperature dependence when the free energy barrier is in the order of $kT$. This supports the idea that freezing becomes a barrierless process around the spinodal temperature.
\end{abstract}


%64.60.Qb	Nucleation
%61.46.Df	Nanoparticles
%64.60.My       Metastable phases
\maketitle
%Introduction
When a liquid is cooled below its freezing temperature we generally expect the system to crystallize. However, the nucleation process requires the formation of a small embryo of the new stable phase that introduces an energetically unfavourable liquid-solid interface and creates a free energy barrier between the liquid and solid phases. As long as the fluctuations in the liquid only result in the formation of embryos smaller than the critical size needed to overcome the barrier, the system will remain fluid. This is the metastable liquid~\cite{pablo_book}.

As the liquid is cooled further, the free energy barrier decreases in height, making nucleation more likely and shortening the lifetime of the metastable liquid. The question then arises: Is there a temperature below which the metastable liquid becomes unstable with respect to all fluctuations? 
Mean-field theories, such as the gradient theory developed by Cahn and Hilliard~\cite{gt}, predict such a limit of stability, or spinodal, for first order phase transitions like the condensation of a gas or liquid-liquid phase separation in a mixture. They also predict that the size of the critical nucleus diverges as the spinodal is approached as a result of the divergence in the mean-field isothermal compressibility of the fluid~\cite{wil04}, and that the nucleation lifetime should diverge, despite the fact that the free energy barrier is in the order of $kT$, as the dynamics become increasingly cooperative~\cite{klein83}. However, recent experiments of phase separating polymers and simulations of the Ising model suggest that the size of the critical embryo remains finite as the spinodal is approached~\cite{pan06}.

Whether a deeply supercooled single component liquid exhibits a spinodal singularity with respect to the crystal remains an open question~\cite{pablo_book}. Trudu et al~\cite{tra06} studied freezing in a bulk Lennard-Jones fluid and found nucleation becomes a spatially diffuse and collective phenomenon when the system is deeply supercooled and suggested this was indicative of the presence of a mean-field spinodal. Recent nucleation experiments on water show nucleation times become extremely short when the liquid is highly compressed, thus defining a practical limit of stability to the liquid state~\cite{water}. These results provide strong but indirect evidence for the existence of a thermodynamic limit of stability for the supercooled liquid state. 

In this letter, we directly locate the limit of stability of the liquid phase by calculating the free energy of the cluster using the largest-sized solid embryo as an order parameter. At temperatures just below freezing, the free energy exhibits a local minimum associated with the metastable liquid and a free energy barrier that separates this liquid from the solid phase. The height of the barrier decreases as the temperature is lowered and eventually disappears so that the free energy becomes a monotonically decreasing function of the order parameter and the liquid phase becomes unstable. This provides the first direct measurement of the spinodal in a simple liquid system.

A rigorous molecular theory of a metastable system requires the introduction of constraints that prevent the system from accessing regions of phase space that will cause the system to evolve towards the more stable state. In the context of a supercooled liquid, this means we need to prevent the appearance of solid-like embryos above the critical size that would cause the liquid to freeze, which suggests we should choose the size of the largest solid-like embryo, $n_{max}$, as an order parameter to describe the state of the cluster~\cite{kline07}. Furthermore, $n_{max}$ seems to be a particularly appropriate order parameter in small nanoscale systems where the nucleation volume is sufficiently small that the appearance of a single post-critical embyro leads to the  termination of the metastable state throughout the entire system. When $n_{max}=0$, the cluster is completely liquid but when $n_{max}=N$, where $N$ is the number of atoms in the cluster, then the cluster is completely solid, as a single crystal. The probability of finding the cluster in a given state is then
\begin{equation} 
P(n_{max})=\frac{Q(n_{max})}{\sum_{n_{max}=0}^{N}Q(n_{max})}\mbox{ ,}
\label{pnmax}
\end{equation}
where 
\begin{equation}
Q(n_{max})=(1/N!\Lambda^{3N})\sum e^{-E(n_{max})/kT}\mbox{ ,}\\
\label{qnmax}
\end{equation}
is the canonical partition function for the system constrained to contain {\it at  least one} largest embryo of size $n_{max}$ so the sum is over all states characterised with a given $n_{max}$,  $E(n_{max})$ is the energy, $k$ is Boltzmann's constant, $T$ is the temperature and $\Lambda$ is the de Broglie wavelength. $P(n_{max})$ can be calculated by simulation and the free energy obtained from
\begin{equation}
\Delta F(n_{max})=-kT\ln P(n_{max})\mbox{ ,}\\
\label{def_fe}
\end{equation}
where $\Delta F(n_{max})$ is the work required to take the entire system from a state where there is no solid-like cluster present, to a state where there is at least one largest cluster of size $n_{max}$. Eq. \ref{def_fe} closely resembles the intensive free energy introduced by ten Wolde and Frenkel~\cite{wolde,colloids} to calculate the free energy barriers associated with nucleation,
\begin{equation}
\Delta F(n)=-kT\ln\left(P_{n}/N\right)\approx -kT\ln\left(N_{n}/N\right)\mbox{ ,}\\
\label{Ffe}
\end{equation}
where $P_{n}$ is the probability of observing an $n-sized$ embryo and $N_{n}$ is the equilibrium number of embryos. $\Delta F(n)$ is the work of forming an $n-sized$ embryo within the metastable phase. In the limit that embryos are rare (i.e. under conditions of mild undercooling) $P(n_{max})$ is approximately equal to the equilibrium number of embryos~\cite{rare} and the two free energies become equivalent  within an additive constant, but it should be stressed that the two free energies are fundamentally different and that we would expect them to behave differently in deeply supercooled systems.

Bagchi et al~\cite{bag_cm} have recently used Eq.~\ref{def_fe} to identify the liquid-gas spinodal in the supersaturated Lennard-Jones gas as the point at which $\Delta F(n_{max})$ is a monotonically decreasing function of $n_{max}$. They find this occurs at a supersaturation consistent with previous estimates of the spinodal~\cite{ljspin} and that the nucleation mechanism in the deeply metastable system changes from classical nucleation, characterised by fluctuating growth of a single large embryo, to a mechanism involving the coalescence of embryos. However, from the definition of $P(n_{max})$, it should be apparent that the free energy is an extensive quantity and it is likely that the location of the spinodal in a bulk system would shift  depending on the number of particles in the simulation. In a small, finite-sized system, such as a liquid nanoparticle, the applicability of Eq.~\ref{def_fe} is clearer.
  
%simulation details
We have previously calculated $\Delta F(n)$ for a gold cluster with $N=456$ atoms, for temperatures above $T=690$K~\cite{ed1}, but that work focused on the role of wetting phenomena and the location of the embryo at the nanoparticle surface. In the present paper, we calculate both $\Delta F(n_{max})$ and $\Delta F(n)$ to lower temperatures for the same cluster in search of the limit of stability of the metastable liquid cluster using the same approach of combining umbrella Monte Carlo (MC) sampling simulation techniques with parallel tempering. We use the semi-empirical embedded-atom method (EAM) potential~\cite{gold_p} to describe the atomic interactions and study the cluster in the $N,V,T$ ensemble with a simulation cell of $V=1500\AA^{3}$ and periodic boundaries. At each temperature, we run eight parallel simulations or windows, each with a parabolic biasing potential $w(n_{max})=0.0005(n_{max}-n_{0})^2$ which biases the system to sample states where the largest embryo, $n_{max}$, in the cluster is around $n_{0}$.  We choose $n_{0}=0,10,20,30\ldots70$  and use $T=750,730,710,690,680,670,660,650$ for tempering. 

Our embryo criterion has been previously described in ref.~\cite{ed1} and closely follows that developed by Frenkel~\cite{colloids} to study crystal nucleation in hard sphere colloids. In brief, the criterion identifies which atoms in the cluster are solid-like by considering the degree to which the local order around two neighbouring atoms is correlated.  If the local order of two atoms is highly correlated, then they are considered to be {\it connected}. If an atom is {\it connected} to half or more of its neighbours, then we consider the atom to be solid-like. Two solid-like atoms are considered to be in the same embryo if they are connected and $n_{max}$ is the largest embryo. 

The embryo criterion is computationally expensive to apply so we use trajectories that consist of 10 normal MC moves per atom sampling the atomic interaction potential, followed by a test against $w(n_{max})$. If the final move is rejected, the system is returned to state at the beginning of the trajectory. We attempt switches between neighboring  $n_{0}$ windows ($T$ fixed) every 10 trajectories. We also attempt switches in neighboring temperatures ($n_{0}$ fixed) every 10 trajectories, but these are offset with the $n_{o}$ switches. These tempering switches have acceptance ratios of about 0.4 and 0.6, respectively. The free energies in each window differ by an unknown additive constant, so the full free energy curve is constructed by fitting the curves to a polynomial in $n_{max}$~\cite{colloids} and a total of $1.74\times 10^6$ trajectories are sampled in each window.

\begin{figure}[t]
\hbox to \hsize{\epsfxsize=0.99\hsize\hfil\epsfbox{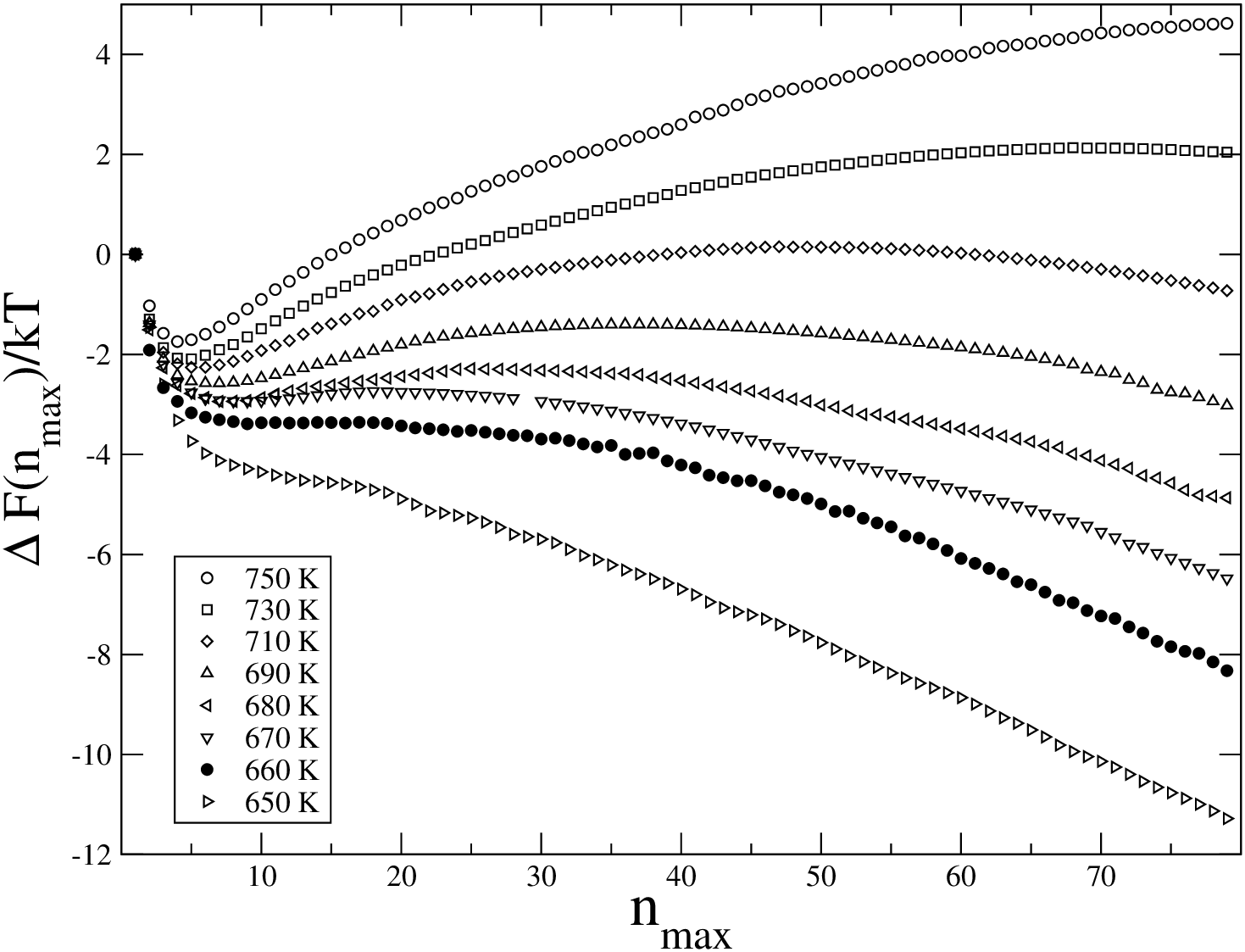}}
\caption{$\Delta F(n_{max})$ as a function of $n_{max}$ for temperatures in the range $T=750-650$.}
\label{fig:free_energy}
\end{figure}

% results+discussion
Fig.~\ref{fig:free_energy} shows the free energy calculated using Eq.~\ref{def_fe}. At temperatures just below the freezing temperature for the cluster, $\Delta F(n_{max})$ exhibits a minimum at small values of $n_{max}$ before it increases to a maximum at larger embryo sizes. $n_{max}=n^{*}_{max}$ denotes the critical embryo size. Fluctuations in the cluster that keep the largest embryo below its critical size are locally stable and represent the configuration space available to the metastable liquid, while larger fluctuations cause the system to freeze. The critical size identifies the constraint required to keep the liquid in metastable equilibrium.

% Pervous studies of nucleation in mean-field system have used a similar approach to define the boundary of the metastable state~\cite{kline07}.

% discussion of number of embryos at this point

%fig: 

As the temperature is lowered, $n^{*}_{max}$ decreases in size and the barrier becomes smaller. Eventually we reach a point, at $T=660$K, where the barrier disappears and all fluctuations which increase the size of the largest cluster lower the free energy, suggesting  we have reached the limit of stability for the fluid phase. Further decreases in $T$ simply increase the thermodynamic driving force towards forming the solid as the free energy curve becomes steeper. 

\begin{figure}[h]
\hbox to \hsize{\epsfxsize=0.90\hsize\hfil\epsfbox{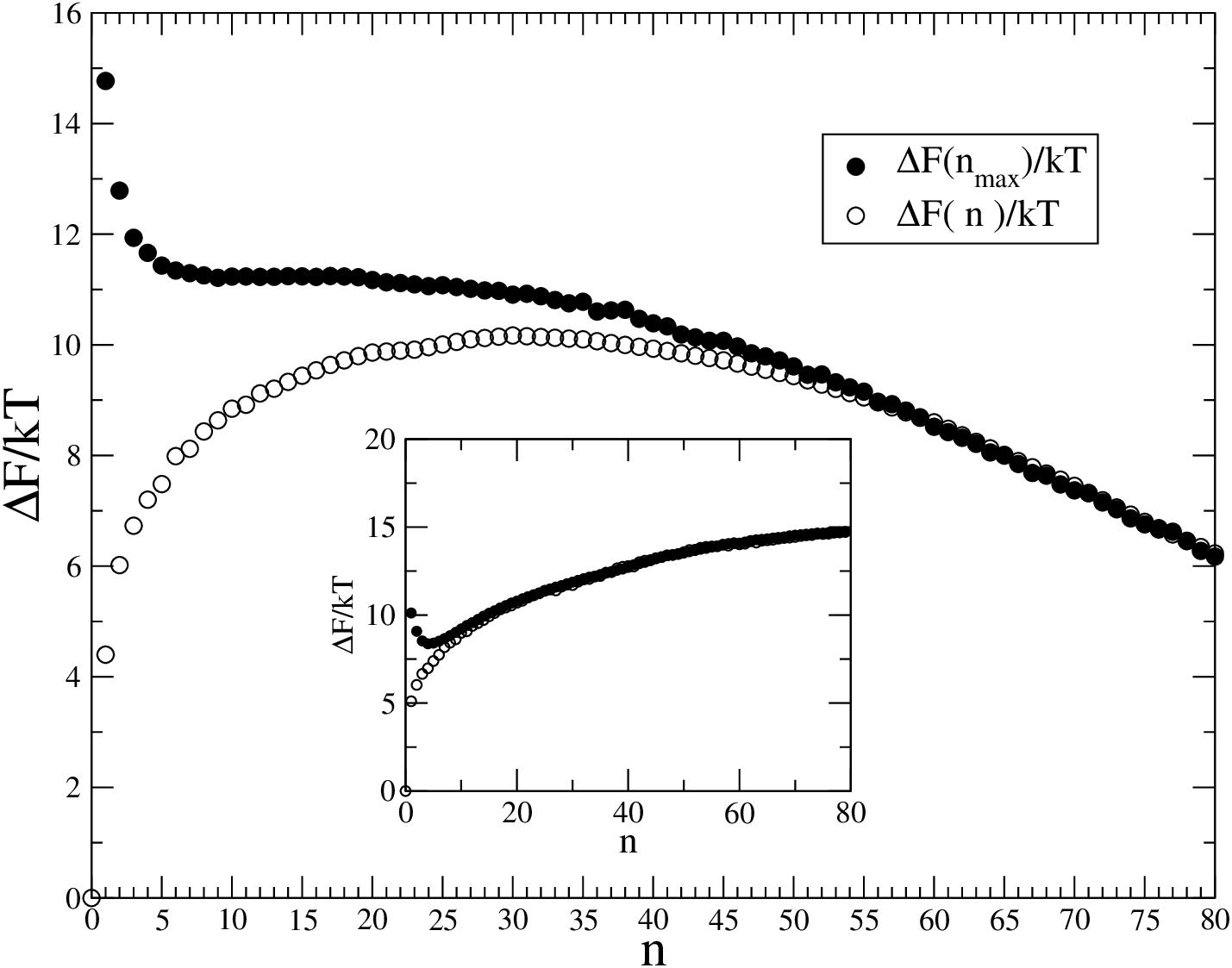}}
\caption{Comparision of $\Delta F(n)$ and $\Delta F(n_{max})$ at $T=660$K and $T=750$K (insert). }
\label{fig:comp}
\end{figure}

%comparison of free energies.
Fig.~\ref{fig:comp} shows the two free energies calculated from Eqs.~\ref{def_fe} and \ref{Ffe} where $\Delta F(n_{max})$ has been shifted vertically to maximise the overlap between the two curves. At $T=750$K (see insert), the two free energies are identical for embryo sizes larger than about 15 since there is generally just one large embryo in the system. The minimum in $\Delta F(n_{max})$ suggests the cluster usually contains a largest cluster of $n\approx 5$, but since $\Delta F(n)$ continues to decrease, there must be a larger number of smaller embryos present. At the spinodal temperature, the two curves are very different and only overlap at the largest embryo sizes. 

%\begin{figure}[h]
%\hbox to \hsize{\epsfxsize=0.90\hsize\hfil\epsfbox{doubleGat750}}
%\vspace{0.2cm}
%\hbox to \hsize{\epsfxsize=0.90\hsize\hfil\epsfbox{doubleGat660}}
%\caption{Comparision of $\Delta F(n)$ and $\Delta F(n_{max})$ at (a) $T=750$K and (b) $T=660$K.}
%\label{fig:comp}
%\end{figure}

\begin{figure}[t]
\hbox to \hsize{\epsfxsize=1\hsize\hfil\epsfbox{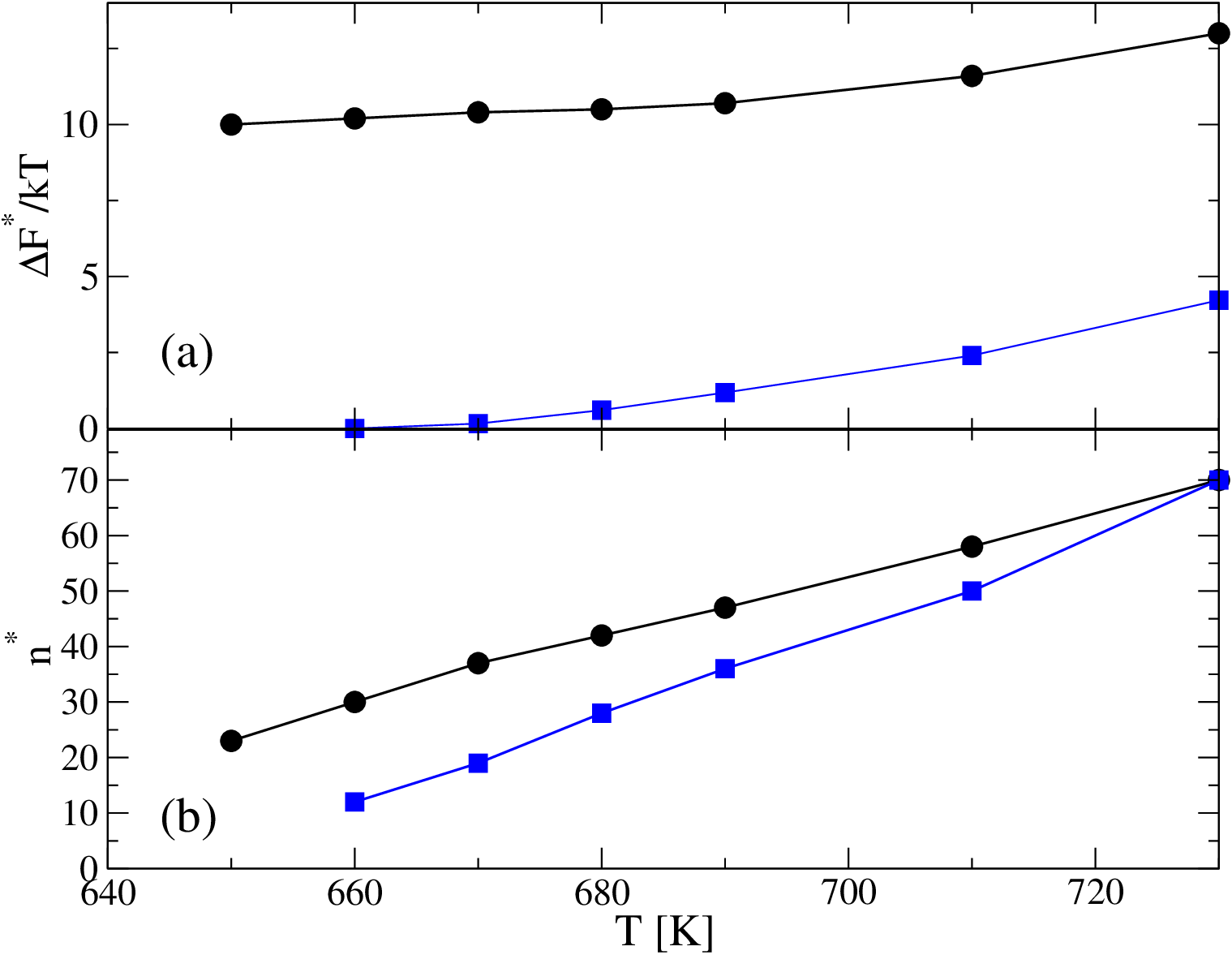}}
\caption{(a) The height of the free energy barrier obtained from the free energy defined in Eq.~\ref{def_fe} (squares) compared to that obtained from a free energy based on the equilibrium distribution of embryos (circles). See ref.~\cite{ed1} for details. (b) The size of the critical embryo obtained using the two methods. Symbols are the same as (a).}
\label{fig:barrier}
\end{figure}

If we define the height of the barrier, $\Delta F(n^{*}_{max})$, as the difference in free energy between the maximum and the small embryo minimum, we can compare this with the usual nucleation barrier, $\Delta F(n^{*})$. Fig.~\ref{fig:barrier}a shows that as the $\Delta F(n^{*}_{max})$ goes to zero at the spinodal, $\Delta F(n^{*})$ plateaus as a function of temperature at around $10kT$. At the same time, the size of the critical embryo for both free energies decreases as a function of temperature. At the spinodal, $\Delta F(n_{max})$ exhibits a flat region, where the embryo sizes in the range $n_{max}=5-25$ have approximately the same free energy, so we can expect considerable fluctuations in the embryo size. Nevertheless, the $n_{max}$ remains finite (Fig.~\ref{fig:barrier}b). This is in direct contrast to the predictions of mean-field theory~\cite{gt,wil04}, but our results are consistent with those of Pan et al~\cite{pan06} and Bagachi et al~\cite{bag_cm}.

% role of cluster criterion

% fig: Compare barriers and critical embryo

% nucleation rate
% presence of glass transition would serve to slow nucleation
The rate at which clusters freeze can be determined by considering an ensemble of temperature quenched, molecular dynamics (MD) simulations~\cite{bart01}. The liquid cluster is initially equilibrated at $T=900$K, well above the freezing temperature, before the temperature is instantaneously quenched below freezing by rescaling the particle velocities. The MD trajectory is then followed as the cluster freezes. Assuming this process is described by a first order rate law, the nucleation rate, $J$, can be obtained from the relation
\begin{equation}
\ln [R(t)]=-JV_{c}(t-t_{0})\mbox{ ,}\\
\label{eq:rate}
\end{equation}
where $R(t)$ is the fraction of un-nucleated clusters at time $t$, $V_{c}$ is the volume of the cluster and $t_{0}$ is the nucleation lag time, which is the time required to reach the steady state concentration of precritical embryos after the quench. To make use of Eq.~\ref{eq:rate}, we consider a cluster to have nucleated when $n_{max}$ is greater than 85 for the last time during the simulation, which runs for 500 picoseconds. The nucleation size is defined as 85 because it is  larger than the critical embryo size at all temperatures studied. A total of 300 quenched simulations are used at each temperature and even at the slowest rates (highest temperatures), less than 5\% of the clusters remained un-nucleated by the end of the simulation. The volume of the cluster was determined using a ``rolling sphere" algorithm~\cite{volc} which defines the surface of a cluster using a hard sphere probe. In our case, the radius of the probe sphere and the gold atoms was taken to be $1.5\AA$. At $T=750$K, $V_{c}=7\times10^3\pm250\AA^{3}$, which is 12\% smaller than would be predicted based on the volume per molecule of bulk liquid EAM gold~\cite{st_eam}. 

\begin{figure}[t]
\hbox to \hsize{\epsfxsize=1\hsize\hfil\epsfbox{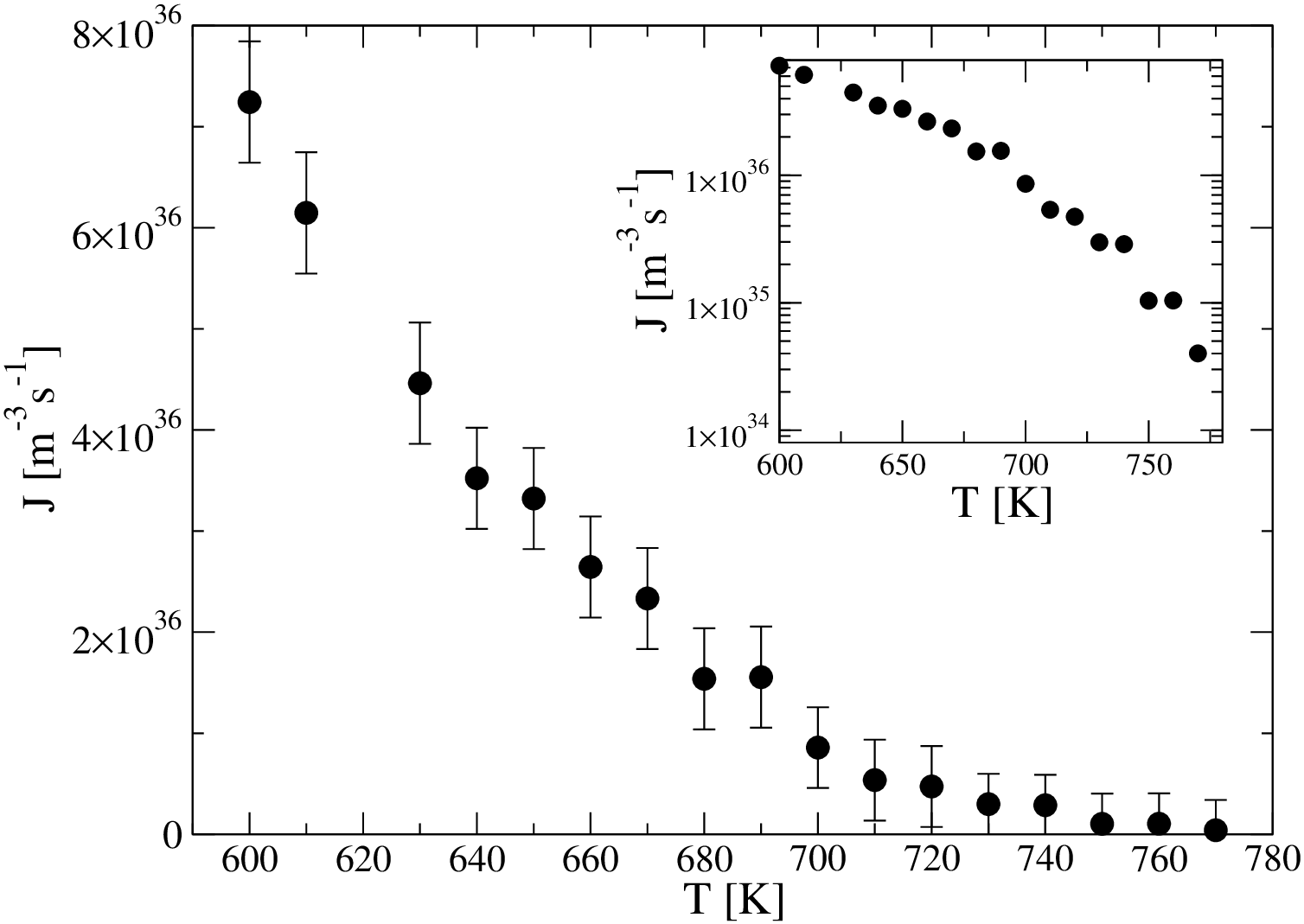}}
\caption{The nucleation rate as a function of temperature. Insert. The same rate data on a log scale.}
\label{fig:rate}
\end{figure}

Fig.~\ref{fig:rate} shows that the nucleation rate increases as the cluster is quenched to lower temperatures. For temperatures below 700K, our rates are approximately the same as those obtained by Bartell et al~\cite{bart01}, who used the same technique, but a larger cluster volume and a different nucleation criterion. Around $T=700$K we see an unexpected increase in the rate with the slope $\partial J/\partial T$ becoming more negative. Classical nucleation theory expresses the rate of nucleation as
\begin{equation}
J=K\exp[-\Delta F^{*}/kT]\mbox{ ,}\\
\label{eq:cnt}
\end{equation}
where the kinetic prefactor is given by $K=24\rho_{n}ZDn^{*2/3}/\lambda$,  $D$ is the diffusion coefficient, $\rho_{n}$ is the number density of particles, $\lambda$ is the typical distance a particle must diffuse in order to join the embryo and $Z=(|\Delta\mu |/6\pi kTn^{*})^{1/2}$ is the Zeldovich factor. $\Delta\mu$ is the difference in chemical potential between the nucleating stable and metastable phases. The temperature dependent parameters in the rate should vary continuously as a function of temperature and cannot account for the rapid increase in rate while Fig.~\ref{fig:barrier}a suggests that the temperature dependence of $\Delta F(n^{*})/kT$ 
would cause the rate to slow, rather than accelerate. However, at $T=700$K, the barrier defined by $\Delta F(n^{*}_{max})/kT$ is in the order of $kT$, which suggests the observed deviation in the temperature dependence of the rate might be associated with a crossover from a barrier dominated nucleation process to a barrierless one. Consequently, both our direct barrier calculations and the independent MD rate calculations point to the strong possibility of a spinodal signifying the limit of stability of the fluid phase.

% fig: rate of nucleation

\acknowledgements
We would like to thank P. H. Poole and S. S. Ashwin for useful discussions.  We acknowledge NSERC for funding and WESTGRID for computing resources.

\end{document}